\documentclass[aps,%
 reprint,
 amsmath,amssymb,
]{revtex4-1}

\usepackage{graphicx}
\usepackage{bm}

\usepackage{algorithm}
\usepackage{algpseudocode}
\usepackage{float}

\begin{document}

\preprint{APS/123-QED}

\title{Controlling the Minimal Feature Sizes in Adjoint Optimization of Nanophotonic Devices Using B-spline Surfaces}

\author{Erfan Khoram$^{1}$, Xiaoping Qian$^{2}$, Ming Yuan$^{3}$, Zongfu Yu$^{1}$}
\altaffiliation{$^{1}$Department of Electrical and Computer Engineering, University of Wisconsin Madison-Madison, WI53706, USA\\$^{2}$Department of Mechanical Engineering, University of Wisconsin Madison-Madison, WI53706, USA\\$^{3}$Department of Statistics, Columbia University,New York, NY10027, USA}

\begin{abstract}
Adjoint optimization is an effective method in the inverse design of nanophotonic devices. In order to ensure the manufacturability, one would like to have control over the minimal feature sizes. Here we propose utilizing a level-set method based on b-spline surfaces in order to control the feature sizes. This approach is first used to design a wavelength demultiplexer. It is also used to implement a nanophotonic structure for artificial neural computing. In both cases, we show that the minimal feature sizes can be easily parameterized and controlled. 

\end{abstract}

\pacs{Valid PACS appear here}
\maketitle

\section*{INTRODUCTION}

Adjoint method has become a very useful tool in the inverse design of photonics\cite{lalau2013adjoint,callewaert2018inverse,piggott2017fabrication,piggott2015inverse,frei2007geometry,su2018fully,pestourie2018inverse,frellsen2016topology,jensen2011topology,michaels2018inverse,hughes2018adjoint,frandsen2016inverse,frei2008optimization, molesky2018inverse, vercruysse2019analytical, Miller2015} and among other optimization methods such as genetic algorithms\cite{offrein1998very}, swarm optimization\cite{mak2016binary}, or nonlinear optimization\cite{shen2015integrated}, adjoint method is particularly efficient as it computes the gradient of a cost function w.r.t. a large number of design parameters.

The inverse design process mostly involves the optimization of the dielectric constant distribution  $\varepsilon_{(r)}$. In a gradient-based approach, $\varepsilon$ is treated as a continuous variable. However, in practice, the permittivity can only take discrete values determined by the types of materials used. To bridge this gap, one can apply the method of level-set or density-based topology optimization. However, even with level-set, or density-based topology optimization to address the issue of discrete material parameters, we still end up with structural features that might be too small for today's nanofabrication. Additional constraints need to be applied to control the size of minimum features. For example Ref\cite{shen2015integrated} defines the medium as a collection of pixels with a fixed size larger than the critical dimension of fabrication; while \cite{hughes2018adjoint} uses the method proposed in \cite{zhou2015minimum}: a spatial low pass filter which controls the feature sizes. Ref \cite{su2017inverse} uses the convolution between the gray-scale medium and a disk that defines the minimum feature size. Refs \cite{frei2007geometry,frei2008optimization} use radial basis functions to control the feature size. In this work, we are going to use b-splines for topology optimization, which has been a highly successful and proven approach for size control in structural shape optimization in the field of mechanics \cite{qian2013topology}.


\section*{B-SPLINES}

Splines are functions generally used for interpolating data and designing curves and surfaces in computer aided designs and computer graphics. These functions are generated by a linear combination of piece-wise polynomial functions. Their strength is in the fact that they are capable of accurate interpolation of data with low degree polynomial components, that would otherwise require a high degree non-local polynomial. The degree of the spline is the degree of its highest order polynomial.

One way of representing a spline is by defining a set of basis functions, called basis-spline functions or b-spline functions. In order to define any one of these basis functions, the domain over which the spline is going to be defined, has to be segmented with a set of \textbf{\textit{knots}} $U = [u_0,...,u_m]$. In this work we only work with knots that appear only once in this set, and are equidistant from one another. Another property that is required to define a b-spline function is its degree (\textbf{\textit{k}}). With both knots and degree defined, basis-spline functions can be expressed using the \textbf{\textit{Cox-de Boor recursion formula}} \cite{de1972calculating,cox1972numerical}.

\begin{equation} \label{eq_1}
\begin{gathered}
N_{(u)}^{i,0}=\left\{
                \begin{array}{ll}
                  1\quad u_i<u<u_{i+1}\\
                  0\quad otherwise
                \end{array}
                \right. \\
N_{(u)}^{i,k}=\frac{u-u_i}{u_{i+k}-u_i}N_{(u)}^{i,k-1}+\frac{u_{i+k+1}-u}{u_{i+k+1}-u_{i+1}}N_{(u)}^{i+1,k-1}
\end{gathered}              
\end{equation}

In the equation above $N_{(x)}^{i,k}$ shows a b-spline of degree $k$. As it can be seen, a b-spline of degree zero is only non-zero on one of the segments of the domain, while higher degrees span more segments of the domain. As a matter of fact, a b-spline of degree $k$ defined over a domain segmented by $m+1$ knots spans $k+1$ one of these intervals ($m+1-k$ such b-splines are defined over the domain). This can be seen in Fig.1 where a domain  has been segmented by 4 knots.

\begin{figure}[ht]
\centering
\includegraphics[scale=1]{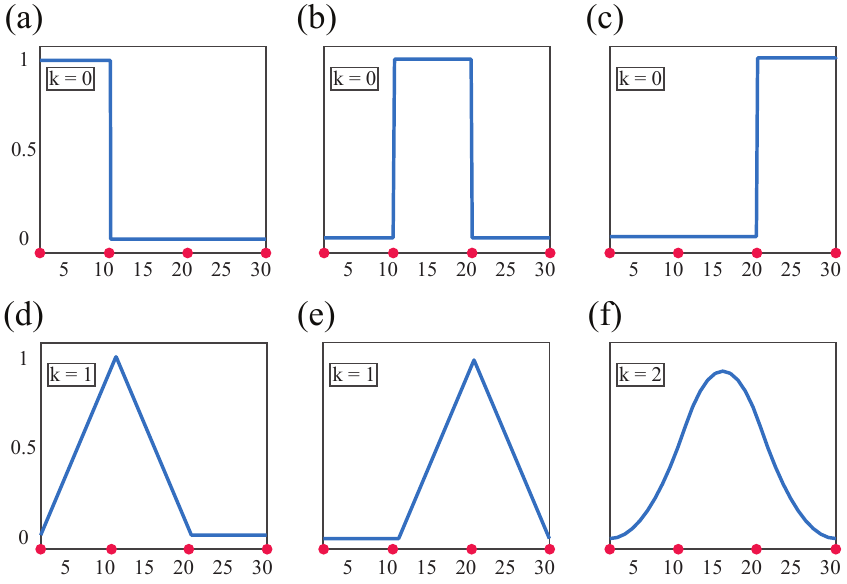}
\caption{A domain of 1 to 31 has been segmented by 4 knots in the set $U=[1,11,21,31]$ (the knots have been have shown with red circles on the x-axis). (a)-(c) show the b-splines of degree 0 each of which span 1 interval. (d) and (e) show the b-splines of degree 1 that span 2 intervals, and (f) shows the b-spline of degree 2 that spans 3 intervals.}
\label{fig:evolution}
\end{figure}

With the basis spline functions, a curve can be generated using a linear combination of these basis functions. In Eq.2, the $P_{(i)}$'s are the set of coefficients that determine the contribution of each b-spline component to the curve.

\begin{equation}\label{eq_2}
    C_{(x)} = \sum_{i=0}^{n_x}N_{(x)}^{i,k}P_{(i)}
\end{equation}

We now discuss the control of feature sizes. The degree of the basis function(k) plays a critical role in controlling the minimal feature size. This can be clearly seen in Fig.1: as the degree of the basis function is increased, the minimum area that it affects also increases and so the minimum size of the features increases. Another way to control the minimal feature size is to tune the spatial density of the knots. A similar effect happens when the knots are distributed more sparsely over the domain. This effect can be seen in Fig.2: the denser the knots are packed together, the smaller the area each of these functions can affect becomes. Consequently, more nuances can be generated in a curve. This is the approach we have taken in this work to control the critical dimensions of our designs.

\begin{figure}[H]
\centering
\includegraphics[scale=1]{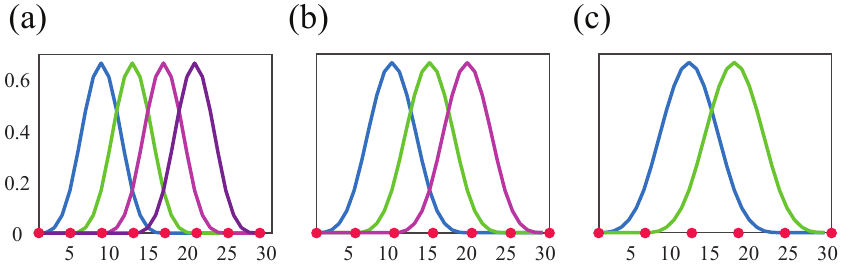}
\caption{The domain from 1 to 31 has been segmented by (a)8, (b)7, and (c)6 knots to create cubic b-splines(b-splines of degree $k=3$). The location of each of the knots has been shown with a red circle on the x-axis.}
\label{fig:evolution}
\end{figure}


\section*{FORMULATION}

B-splines can be used for a density based approach or a  level-set one. Next we use 2 dimensional (2D) photonic design to illustrate this method in the context of the level-set method. In 2D, we use basis spline functions to construct a surface $\phi_{(x,y)}$ using the following equation:

\begin{equation}\label{eq_3}
\begin{gathered}
\phi_{(x,y)} = \sum_{i=0}^{n_x}\sum_{j=0}^{n_y}N_{(x)}^{i,k_x}N_{(y)}^{j,k_y}P_{(i,j)}
\end{gathered}
\end{equation}

Similar to Eq.2, $N_{(x)}^{i,k_x}$ and $N_{(y)}^{j,k_y}$ are basis-spline functions of degree $k_x$ and $k_y$ respectively, and $P_{(i,j)}$ are the set of coefficients. If we were to limit the value of the coefficients between 0 and 1, the generated surface could be used in a density-based topology optimization. Here, we focus on the level-set method.

In order to use the level-set method for topology optimization, we first need to define a surface $\phi_{(x,y)}$ over the area that the optimized device will occupy (The value of each point on this surface can vary across a limited range around zero). Then the medium can be shaped according to Eq.1 based on this surface.

\begin{equation} \label{eq_4}
\varepsilon(r)=\left\{
                \begin{array}{ll}
                  \varepsilon_{1}\quad\phi(r)<0\\
                  \varepsilon_{2}\quad\phi(r)>0
                \end{array}
              \right.
\end{equation}

Now, optimizing the medium with a level-set method consists of evolving the boundaries between the two materials (all the points where $\phi_{(r)}=0$) in a manner that improves the performance of the device. With a b-spline surface for this task, we have direct control on this surface, and consequently on the boundary between the constituent materials. Having defined the level-set surface as a b-spline surface, we can now directly optimize a cost function \textit{J} with respect to the control parameters of the b-spline surface.

\begin{equation} \label{eq_5}
\begin{gathered}
\frac{\partial J}{\partial P_{(i,j)}}=\int_s (\frac{\partial J}{\partial \varepsilon_{(\phi_{(x,y)})}})(\frac{\partial \varepsilon_{ (\phi_{(x,y)}) } }{ \partial \phi_{(x,y)} }) (\frac{\partial \phi_{(x,y)}}{\partial_{(i,j)}})\\
\Rightarrow \frac{\partial J}{\partial P_{(i,j)}}=\int_s (\frac{\partial J}{\partial \varepsilon_{(\phi_{(x,y)})}})  (\delta_{\phi_{(x,y)}})(N_{(x)}^{i,k_x}N_{(y)}^{j,k_y})
\end{gathered}              
\end{equation}

In Eq.5, the first term can be calculated with the adjoint method, the second term(a Dirac delta function which is nonzero only where $\phi_{(x,y)}$ passes zero),can be computed with marching squares algorithm\cite{ISOCONTOUR}, and the final term is a simple matrix multiplication.

Once the gradient of the cost function is calculated, gradient descent can be used to update the b-spline coefficients. However, in order to ensure the stability of the process, in each update iteration the coefficients are normalized by the highest value among them\cite{bernard2009variational}. The mentioned steps can be summarized as shown in Eq.6, where $P^t$ is the matrix of all the b-spline coefficients at iteration $t$ and $||P^t||_\infty$ is the maximum absolute value in this matrix.

\begin{equation} \label{eq_4}
\begin{gathered}
P^{t+1}=P^t-\gamma \frac{\partial J}{\partial P}\Big\rvert_{P=P^t}\\
P^{t+1} = \frac{P^{t+1}}{||P^{t+1}||_\infty}
\end{gathered}              
\end{equation}


\section*{APPLICATION}

In order to demonstrate the application of this approach, we utilized it to design two nanophotonic devices: a wavelength demultiplexer \cite{piggott2015inverse} and a nanophotonic structure for artificial neural computing \cite{khoram2019nanophotonic}.

\textbf{The wavelength demultiplexer} is a three port device that guides the light coming from the input port to one of the two output ports based on its wavelength. In our design, this device is of size $4\,\mu m\times2\,\mu m$ designed to demultiplex $1500\,nm$ and $1550\,nm$ wavelengths. We optimize three versions of this device with the intention of showcasing how the critical dimensions of the device can be tuned by the basis spline functions. 

The b-spline surface is defined on the same grid as the one used for the electromagnetic simulation and over the area to be optimized. We can use either the distribution of the knots on the grid, or the degree of the b-splines to control the critical dimension. In this work we opt for the former. We define our basis functions as cubic b-splines, and use three different knot distributions to realize three version of the devices with different feature sizes. The first design has the knots placed at every second point on the simulation grid, the second design at every fifth point, and the third design at every ninth point. This effectively provides us with 4559, 629, and 152 b-spline coefficients respectively. Once the b-spline surface is set up, we use Eq.4 to define the medium, where $\varepsilon_1$ is set to the permittivity of $SiO_2$ and $\varepsilon_2$ is set to that of $Si$. Finally, a cost function is necessary to optimize the structure. The cost function we define here tries to maximize the the transmission of the time-averaged energy flux to the output port corresponding to the input wavelength.

\begin{figure}[h]
\centering
\includegraphics[scale=1]{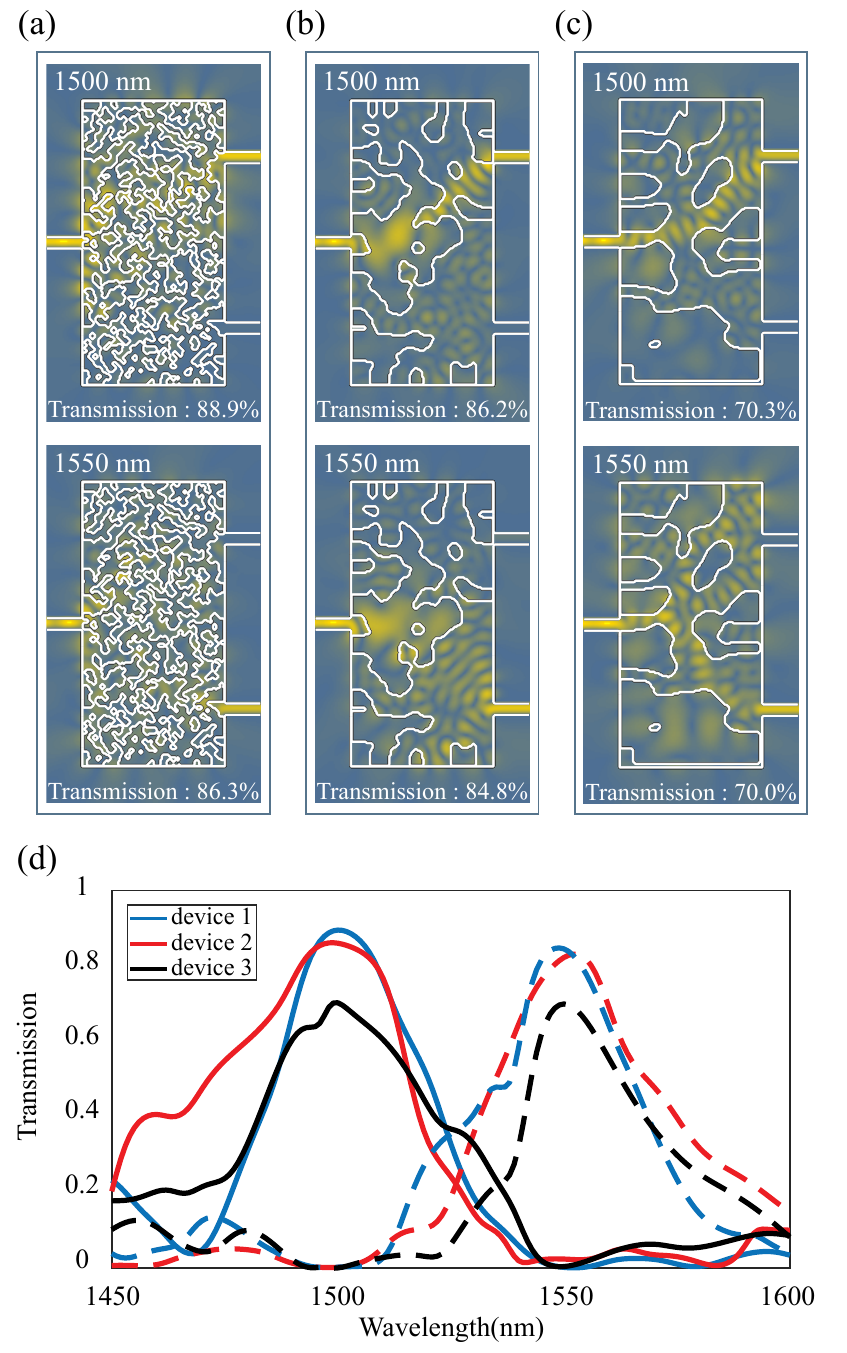}
\caption{The wavelength demultiplexers acting on $1500\,nm$ and $1550\,nm$ wavelengths. These devices are designed with P matrices (matrices of the b-spline coefficients) with sizes (a) $97\times47$, (b) $37\times17$, and (c) $19\times8$. As shown here, the higher the number of coefficients(corresponding to packing the b-spline knots closer together), the smaller the details of the designed device, and the higher the confinement of the field to the device. (d) shows the transmission for each of these three instances. Transmission for $1500\,nm$ is shown with continuous lines, while the dashed lines represent the transmission for $1550\,nm$. The degrading effect of reducing the number of b-spline coefficients on transmission can be clearly seen here.}
\label{fig:evolution}
\end{figure}

The devices were optimized with respect to the cost function mentioned above using Eq.4 and Eq.5. The results for the optimization of these three instances have been depicted in Fig.3. As it was mentioned earlier, since placing the knots more sparsely results in an increase in the area that each basis-spline function affects, the critical dimension of the device increases. This can clearly be seen in Fig.3. As we move from Fig.3(a) to Fig.3(c) the feature sizes distinctly expand. However, even for the case of the sparsely distributed knots, the occasional nuances in the b-spline surface can cause the appearance of small features in the device. This problem is mitigated by applying erosion and dilation processes on the device in each update iteration to eliminate these occasional smaller features. We should point out that this control over the feature sizes of the device comes at a price: there is a trade off between the performance of the device and the critical dimensions of the structure. As we use larger critical dimensions, the  degree of freedom, i.e. the number of b-Spline functions used, decreases, and the optimization is further constrained, consequently the performance generally degrades. This is also evident in Fig.3, where the transmission drops moving from the first device to the last one.

\textbf{Nanophotonic artificial neural computing} This device can accomplish the same task of neural inference as in \cite{khoram2019nanophotonic} . The objective is to recognize the value of a handwritten digit 0-9. The input light from the handwritten digit is focused by a nanophotonic structure to different locations according to the value of the digit. The medium then has to map all the different writing styles of the same digit to the specific location corresponding to that digit; furthermore, the medium must perform well on handwritten digits that it has not yet seen.

The setup for this task is as follows: the image of the handwritten digit is first vectorized and mapped to an array of light sources placed on the left side of the nanophotonic structure. Then a set of ten receivers corresponding to the each of the ten classes 0-9 are placed on the right side of the medium. Inside the device, the input light scatters and reflects off of the numerous material interfaces, and then focuses on a spatial location on the output side. On the output side, the receivers measure the light intensity at their respective locations and the class with the maximum measured value is selected as the correct class for the input image.

The device is made possible by a much more extensive optimization process than that for the wavelength demultiplexer. This optimization process is equivalent to the training process in digital neural networks. We define a cost function and use adjoint method to calculate the gradient. At the time of training, we measure the intensity of light at the location of all the receivers. Then we use a cross-entropy cost function between the normalized output array and a $1\times10$ one-hot vector(a vector that is all zeros except at the index of the correct class), to calculate the cost value for that specific instance. Details of the working principle can be found in Ref\cite{khoram2019nanophotonic}. In this work we implement the level-set function based on a b-spline surface to control the minimal critical dimension. In our previous work \cite{khoram2019nanophotonic}, the level-set method was implemented based on a surface set as a signed distance function. Here, the initial medium is created by generating a random set of values for the b-spline coefficients, and then the medium is created based on the generated surface with the surface evolving afterwards by updating the b-spline coefficients. On the other hand, in Ref\cite{khoram2019nanophotonic} the medium was generated by randomly distributing inclusions made up of a second material inside the host medium, and then a surface $\phi_{(r)}$ is generated based on this initial medium (similar to the previous section, the surface relates to the medium with Eq.4). The structure then evolves by updating this signed-distance surface at each training iteration using a modified version of the following equation.

\begin{equation} \label{eq_s26}
\partial_t\phi+v(x,y)|\nabla\phi|=0
\end{equation}

In this equation $v(x,y)$ is the velocity with which each point on the zero crossing curves of the level-set function moves normal to each of those curves. This velocity is set equal to the gradient acquired with the adjoint method. This implementation of the level-set method for the image classification task turned out to be quite dependant on the initial distribution of the inclusions, and therefore we had to initialize the medium with many small inclusions to get a good performance from the device. Doing so resulted in small feature sizes in the device which make the fabrication of the device rather difficult. However, with the b-spline approach to level-set function this problem does not happen as the optimization process is less dependant on the initial shape of the medium.

\begin{figure}[ht]
\centering
\includegraphics[scale=1]{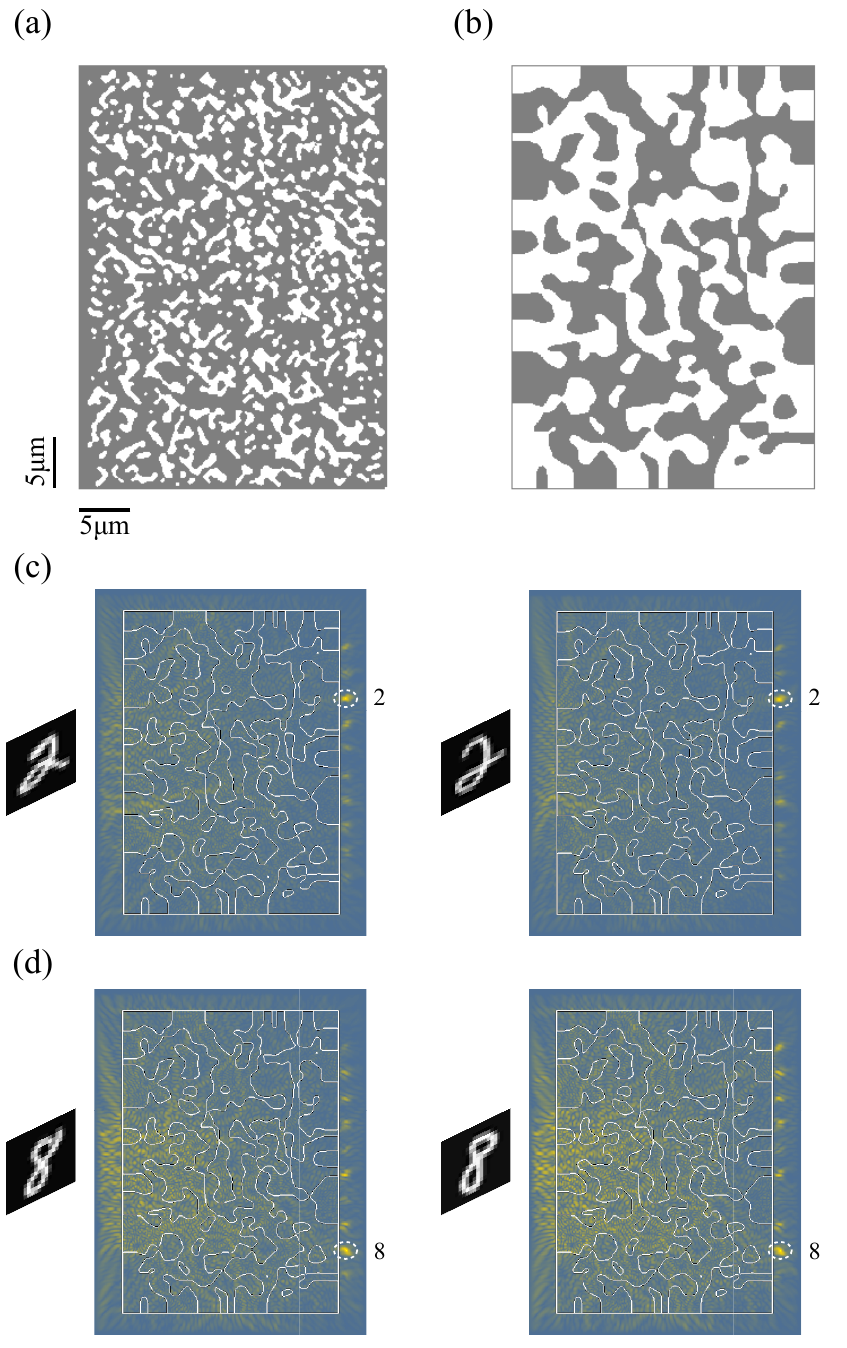}
\caption{Nanophotonic structures can perform artificial computing. The structures are optimized by adjoint method. The two structures can realize similar performance levels but  the b-spine method (b) produces structures that are much easier to fabricate. (a) The structure acquired by following the evolution of a signed distance level-set surface. (b) A medium achieved by following the b-spline surface scheme  with larger minimal feature sizes than (a). (c) Two image samples of the same digit 2 generate different field distributions but they are both recognized as the same digit. (d) The same phenomenon can be seen for two image samples of handwritten digit 8, where different field distributions result in recognition of the same digit.}
\label{fig:evolution}
\end{figure}

We optimize two versions of this device, one with the signed distance function approach, and the other with the b-spline approach. Both devices were optimized for the wavelength $1\,\mu m$ with the size $42\,\mu m\times30\,\mu m$ and made up of $SiO_2$ as the host material and $air$ as the inclusions. For the signed-distance implementation 2000 inclusions of size $5\,\mu m\times5\,\mu m$ are spread throughout the device for initialization, and for the b-spline implementation the knots are distributed at every tenth point on the simulation grid in both axial dimensions. The accuracy on the test set for the two devices turned out to be $76.8\%$ and $77.2\%$. The performance of the b-spline approach is slightly better in this case which is not of great consequence as the gap between the two values can be reduced with better initialization of the signed-distance approach. However, the important point that is apparent in Fig.4 is the difference between the shape of the two structures. As it can be seen, the first device shown in Fig.4(a) has much smaller features whereas the second device depicted in Fig.4(b) has a much larger critical dimension. This makes the second device much easier to fabricate, and thus the b-spline approach much more desirable. Finally, Fig4.(c) and Fig.4(d) show the device in action. As it can be seen, light from different class of images is focused at distinct locations corresponding to that class. Moreover, different image samples of the same digit produce different field distributions; however, these different field distributions result in the same output class.


\section*{CONCLUSION}

In this work, we adopt a proven technique to size control that is used in the field of mechanics to the inverse design of photonics. It use B-spline basis to describe the distribution of the dielectric constant. The minimal feature size is controlled by the knots distribution and the degree. This method is very easy to implement and interpret. We demonstrated the proposed method for a wavelength demultiplexer, and a nanophotonic structure for artificial neural computing. The first class of nanophotonic devices we implemented was used to demonstrate how to handle the trade off between the complexity and their performance. For the second class, two variations of the level-set methods were used to achieve the same goal. The comparison between the two cases shows how having a well-defined set of control parameters can help us in designing a nanophotonic device in one optimization stage, without the need to have a good initial guess of the variables we are trying to optimize.

\section{Acknowledgements}
The authors thank W. Shin for his help on improving the computational speed of the implementation of this method. This work was partially supported by the National Science Foundation under Grant No. 1561917. It was also partially supported by the Defense Advanced Research Projects Agency (DARPA), under agreement HR00111820046. The views, opinions and/or findings expressed are those of the author and should not be interpreted as representing the official views or policies of the Department of Defense or the U.S. Government.

\bibliography{bibl}

\begin{thebibliography}{27}%
\makeatletter
\providecommand \@ifxundefined [1]{%
 \@ifx{#1\undefined}
}%
\providecommand \@ifnum [1]{%
 \ifnum #1\expandafter \@firstoftwo
 \else \expandafter \@secondoftwo
 \fi
}%
\providecommand \@ifx [1]{%
 \ifx #1\expandafter \@firstoftwo
 \else \expandafter \@secondoftwo
 \fi
}%
\providecommand \natexlab [1]{#1}%
\providecommand \enquote  [1]{``#1''}%
\providecommand \bibnamefont  [1]{#1}%
\providecommand \bibfnamefont [1]{#1}%
\providecommand \citenamefont [1]{#1}%
\providecommand \href@noop [0]{\@secondoftwo}%
\providecommand \href [0]{\begingroup \@sanitize@url \@href}%
\providecommand \@href[1]{\@@startlink{#1}\@@href}%
\providecommand \@@href[1]{\endgroup#1\@@endlink}%
\providecommand \@sanitize@url [0]{\catcode `\\12\catcode `\$12\catcode
  `\&12\catcode `\#12\catcode `\^12\catcode `\_12\catcode `\%12\relax}%
\providecommand \@@startlink[1]{}%
\providecommand \@@endlink[0]{}%
\providecommand \url  [0]{\begingroup\@sanitize@url \@url }%
\providecommand \@url [1]{\endgroup\@href {#1}{\urlprefix }}%
\providecommand \urlprefix  [0]{URL }%
\providecommand \Eprint [0]{\href }%
\providecommand \doibase [0]{http://dx.doi.org/}%
\providecommand \selectlanguage [0]{\@gobble}%
\providecommand \bibinfo  [0]{\@secondoftwo}%
\providecommand \bibfield  [0]{\@secondoftwo}%
\providecommand \translation [1]{[#1]}%
\providecommand \BibitemOpen [0]{}%
\providecommand \bibitemStop [0]{}%
\providecommand \bibitemNoStop [0]{.\EOS\space}%
\providecommand \EOS [0]{\spacefactor3000\relax}%
\providecommand \BibitemShut  [1]{\csname bibitem#1\endcsname}%
\let\auto@bib@innerbib\@empty
\bibitem [{\citenamefont {Lalau-Keraly}\ \emph {et~al.}(2013)\citenamefont
  {Lalau-Keraly}, \citenamefont {Bhargava}, \citenamefont {Miller},\ and\
  \citenamefont {Yablonovitch}}]{lalau2013adjoint}%
  \BibitemOpen
  \bibfield  {author} {\bibinfo {author} {\bibfnamefont {C.~M.}\ \bibnamefont
  {Lalau-Keraly}}, \bibinfo {author} {\bibfnamefont {S.}~\bibnamefont
  {Bhargava}}, \bibinfo {author} {\bibfnamefont {O.~D.}\ \bibnamefont
  {Miller}}, \ and\ \bibinfo {author} {\bibfnamefont {E.}~\bibnamefont
  {Yablonovitch}},\ }\href@noop {} {\bibfield  {journal} {\bibinfo  {journal}
  {Optics express}\ }\textbf {\bibinfo {volume} {21}},\ \bibinfo {pages}
  {21693} (\bibinfo {year} {2013})}\BibitemShut {NoStop}%
\bibitem [{\citenamefont {Callewaert}\ \emph {et~al.}(2018)\citenamefont
  {Callewaert}, \citenamefont {Velev}, \citenamefont {Kumar}, \citenamefont
  {Sahakian},\ and\ \citenamefont {Aydin}}]{callewaert2018inverse}%
  \BibitemOpen
  \bibfield  {author} {\bibinfo {author} {\bibfnamefont {F.}~\bibnamefont
  {Callewaert}}, \bibinfo {author} {\bibfnamefont {V.}~\bibnamefont {Velev}},
  \bibinfo {author} {\bibfnamefont {P.}~\bibnamefont {Kumar}}, \bibinfo
  {author} {\bibfnamefont {A.}~\bibnamefont {Sahakian}}, \ and\ \bibinfo
  {author} {\bibfnamefont {K.}~\bibnamefont {Aydin}},\ }\href@noop {}
  {\bibfield  {journal} {\bibinfo  {journal} {Scientific reports}\ }\textbf
  {\bibinfo {volume} {8}},\ \bibinfo {pages} {1358} (\bibinfo {year}
  {2018})}\BibitemShut {NoStop}%
\bibitem [{\citenamefont {Piggott}\ \emph {et~al.}(2017)\citenamefont
  {Piggott}, \citenamefont {Petykiewicz}, \citenamefont {Su},\ and\
  \citenamefont {Vu{\v{c}}kovi{\'c}}}]{piggott2017fabrication}%
  \BibitemOpen
  \bibfield  {author} {\bibinfo {author} {\bibfnamefont {A.~Y.}\ \bibnamefont
  {Piggott}}, \bibinfo {author} {\bibfnamefont {J.}~\bibnamefont
  {Petykiewicz}}, \bibinfo {author} {\bibfnamefont {L.}~\bibnamefont {Su}}, \
  and\ \bibinfo {author} {\bibfnamefont {J.}~\bibnamefont
  {Vu{\v{c}}kovi{\'c}}},\ }\href@noop {} {\bibfield  {journal} {\bibinfo
  {journal} {Scientific Reports}\ }\textbf {\bibinfo {volume} {7}},\ \bibinfo
  {pages} {1786} (\bibinfo {year} {2017})}\BibitemShut {NoStop}%
\bibitem [{\citenamefont {Piggott}\ \emph {et~al.}(2015)\citenamefont
  {Piggott}, \citenamefont {Lu}, \citenamefont {Lagoudakis}, \citenamefont
  {Petykiewicz}, \citenamefont {Babinec},\ and\ \citenamefont
  {Vu{\v{c}}kovi{\'c}}}]{piggott2015inverse}%
  \BibitemOpen
  \bibfield  {author} {\bibinfo {author} {\bibfnamefont {A.~Y.}\ \bibnamefont
  {Piggott}}, \bibinfo {author} {\bibfnamefont {J.}~\bibnamefont {Lu}},
  \bibinfo {author} {\bibfnamefont {K.~G.}\ \bibnamefont {Lagoudakis}},
  \bibinfo {author} {\bibfnamefont {J.}~\bibnamefont {Petykiewicz}}, \bibinfo
  {author} {\bibfnamefont {T.~M.}\ \bibnamefont {Babinec}}, \ and\ \bibinfo
  {author} {\bibfnamefont {J.}~\bibnamefont {Vu{\v{c}}kovi{\'c}}},\ }\href@noop
  {} {\bibfield  {journal} {\bibinfo  {journal} {Nature Photonics}\ }\textbf
  {\bibinfo {volume} {9}},\ \bibinfo {pages} {374} (\bibinfo {year}
  {2015})}\BibitemShut {NoStop}%
\bibitem [{\citenamefont {Frei}\ \emph {et~al.}(2007)\citenamefont {Frei},
  \citenamefont {Tortorelli},\ and\ \citenamefont
  {Johnson}}]{frei2007geometry}%
  \BibitemOpen
  \bibfield  {author} {\bibinfo {author} {\bibfnamefont {W.}~\bibnamefont
  {Frei}}, \bibinfo {author} {\bibfnamefont {D.}~\bibnamefont {Tortorelli}}, \
  and\ \bibinfo {author} {\bibfnamefont {H.}~\bibnamefont {Johnson}},\
  }\href@noop {} {\bibfield  {journal} {\bibinfo  {journal} {Optics letters}\
  }\textbf {\bibinfo {volume} {32}},\ \bibinfo {pages} {77} (\bibinfo {year}
  {2007})}\BibitemShut {NoStop}%
\bibitem [{\citenamefont {Su}\ \emph {et~al.}(2018)\citenamefont {Su},
  \citenamefont {Trivedi}, \citenamefont {Sapra}, \citenamefont {Piggott},
  \citenamefont {Vercruysse},\ and\ \citenamefont
  {Vu{\v{c}}kovi{\'c}}}]{su2018fully}%
  \BibitemOpen
  \bibfield  {author} {\bibinfo {author} {\bibfnamefont {L.}~\bibnamefont
  {Su}}, \bibinfo {author} {\bibfnamefont {R.}~\bibnamefont {Trivedi}},
  \bibinfo {author} {\bibfnamefont {N.~V.}\ \bibnamefont {Sapra}}, \bibinfo
  {author} {\bibfnamefont {A.~Y.}\ \bibnamefont {Piggott}}, \bibinfo {author}
  {\bibfnamefont {D.}~\bibnamefont {Vercruysse}}, \ and\ \bibinfo {author}
  {\bibfnamefont {J.}~\bibnamefont {Vu{\v{c}}kovi{\'c}}},\ }\href@noop {}
  {\bibfield  {journal} {\bibinfo  {journal} {Optics express}\ }\textbf
  {\bibinfo {volume} {26}},\ \bibinfo {pages} {4023} (\bibinfo {year}
  {2018})}\BibitemShut {NoStop}%
\bibitem [{\citenamefont {Pestourie}\ \emph {et~al.}(2018)\citenamefont
  {Pestourie}, \citenamefont {P{\'e}rez-Arancibia}, \citenamefont {Lin},
  \citenamefont {Shin}, \citenamefont {Capasso},\ and\ \citenamefont
  {Johnson}}]{pestourie2018inverse}%
  \BibitemOpen
  \bibfield  {author} {\bibinfo {author} {\bibfnamefont {R.}~\bibnamefont
  {Pestourie}}, \bibinfo {author} {\bibfnamefont {C.}~\bibnamefont
  {P{\'e}rez-Arancibia}}, \bibinfo {author} {\bibfnamefont {Z.}~\bibnamefont
  {Lin}}, \bibinfo {author} {\bibfnamefont {W.}~\bibnamefont {Shin}}, \bibinfo
  {author} {\bibfnamefont {F.}~\bibnamefont {Capasso}}, \ and\ \bibinfo
  {author} {\bibfnamefont {S.~G.}\ \bibnamefont {Johnson}},\ }\href@noop {}
  {\bibfield  {journal} {\bibinfo  {journal} {Optics express}\ }\textbf
  {\bibinfo {volume} {26}},\ \bibinfo {pages} {33732} (\bibinfo {year}
  {2018})}\BibitemShut {NoStop}%
\bibitem [{\citenamefont {Frellsen}\ \emph {et~al.}(2016)\citenamefont
  {Frellsen}, \citenamefont {Ding}, \citenamefont {Sigmund},\ and\
  \citenamefont {Frandsen}}]{frellsen2016topology}%
  \BibitemOpen
  \bibfield  {author} {\bibinfo {author} {\bibfnamefont {L.~F.}\ \bibnamefont
  {Frellsen}}, \bibinfo {author} {\bibfnamefont {Y.}~\bibnamefont {Ding}},
  \bibinfo {author} {\bibfnamefont {O.}~\bibnamefont {Sigmund}}, \ and\
  \bibinfo {author} {\bibfnamefont {L.~H.}\ \bibnamefont {Frandsen}},\
  }\href@noop {} {\bibfield  {journal} {\bibinfo  {journal} {Optics express}\
  }\textbf {\bibinfo {volume} {24}},\ \bibinfo {pages} {16866} (\bibinfo {year}
  {2016})}\BibitemShut {NoStop}%
\bibitem [{\citenamefont {Jensen}\ and\ \citenamefont
  {Sigmund}(2011)}]{jensen2011topology}%
  \BibitemOpen
  \bibfield  {author} {\bibinfo {author} {\bibfnamefont {J.~S.}\ \bibnamefont
  {Jensen}}\ and\ \bibinfo {author} {\bibfnamefont {O.}~\bibnamefont
  {Sigmund}},\ }\href@noop {} {\bibfield  {journal} {\bibinfo  {journal} {Laser
  \& Photonics Reviews}\ }\textbf {\bibinfo {volume} {5}},\ \bibinfo {pages}
  {308} (\bibinfo {year} {2011})}\BibitemShut {NoStop}%
\bibitem [{\citenamefont {Michaels}\ and\ \citenamefont
  {Yablonovitch}(2018)}]{michaels2018inverse}%
  \BibitemOpen
  \bibfield  {author} {\bibinfo {author} {\bibfnamefont {A.}~\bibnamefont
  {Michaels}}\ and\ \bibinfo {author} {\bibfnamefont {E.}~\bibnamefont
  {Yablonovitch}},\ }\href@noop {} {\bibfield  {journal} {\bibinfo  {journal}
  {Optics express}\ }\textbf {\bibinfo {volume} {26}},\ \bibinfo {pages} {4766}
  (\bibinfo {year} {2018})}\BibitemShut {NoStop}%
\bibitem [{\citenamefont {Hughes}\ \emph {et~al.}(2018)\citenamefont {Hughes},
  \citenamefont {Minkov}, \citenamefont {Williamson},\ and\ \citenamefont
  {Fan}}]{hughes2018adjoint}%
  \BibitemOpen
  \bibfield  {author} {\bibinfo {author} {\bibfnamefont {T.~W.}\ \bibnamefont
  {Hughes}}, \bibinfo {author} {\bibfnamefont {M.}~\bibnamefont {Minkov}},
  \bibinfo {author} {\bibfnamefont {I.~A.}\ \bibnamefont {Williamson}}, \ and\
  \bibinfo {author} {\bibfnamefont {S.}~\bibnamefont {Fan}},\ }\href@noop {}
  {\bibfield  {journal} {\bibinfo  {journal} {ACS Photonics}\ }\textbf
  {\bibinfo {volume} {5}},\ \bibinfo {pages} {4781} (\bibinfo {year}
  {2018})}\BibitemShut {NoStop}%
\bibitem [{\citenamefont {Frandsen}\ and\ \citenamefont
  {Sigmund}(2016)}]{frandsen2016inverse}%
  \BibitemOpen
  \bibfield  {author} {\bibinfo {author} {\bibfnamefont {L.~H.}\ \bibnamefont
  {Frandsen}}\ and\ \bibinfo {author} {\bibfnamefont {O.}~\bibnamefont
  {Sigmund}},\ }in\ \href@noop {} {\emph {\bibinfo {booktitle} {Photonic and
  Phononic Properties of Engineered Nanostructures VI}}},\ Vol.\ \bibinfo
  {volume} {9756}\ (\bibinfo {organization} {International Society for Optics
  and Photonics},\ \bibinfo {year} {2016})\ p.\ \bibinfo {pages}
  {97560Y}\BibitemShut {NoStop}%
\bibitem [{\citenamefont {Frei}\ \emph {et~al.}(2008)\citenamefont {Frei},
  \citenamefont {Johnson},\ and\ \citenamefont
  {Choquette}}]{frei2008optimization}%
  \BibitemOpen
  \bibfield  {author} {\bibinfo {author} {\bibfnamefont {W.~R.}\ \bibnamefont
  {Frei}}, \bibinfo {author} {\bibfnamefont {H.}~\bibnamefont {Johnson}}, \
  and\ \bibinfo {author} {\bibfnamefont {K.~D.}\ \bibnamefont {Choquette}},\
  }\href@noop {} {\bibfield  {journal} {\bibinfo  {journal} {Journal of Applied
  Physics}\ }\textbf {\bibinfo {volume} {103}},\ \bibinfo {pages} {033102}
  (\bibinfo {year} {2008})}\BibitemShut {NoStop}%
\bibitem [{\citenamefont {Molesky}\ \emph {et~al.}(2018)\citenamefont
  {Molesky}, \citenamefont {Lin}, \citenamefont {Piggott}, \citenamefont {Jin},
  \citenamefont {Vuckovi{\'c}},\ and\ \citenamefont
  {Rodriguez}}]{molesky2018inverse}%
  \BibitemOpen
  \bibfield  {author} {\bibinfo {author} {\bibfnamefont {S.}~\bibnamefont
  {Molesky}}, \bibinfo {author} {\bibfnamefont {Z.}~\bibnamefont {Lin}},
  \bibinfo {author} {\bibfnamefont {A.~Y.}\ \bibnamefont {Piggott}}, \bibinfo
  {author} {\bibfnamefont {W.}~\bibnamefont {Jin}}, \bibinfo {author}
  {\bibfnamefont {J.}~\bibnamefont {Vuckovi{\'c}}}, \ and\ \bibinfo {author}
  {\bibfnamefont {A.~W.}\ \bibnamefont {Rodriguez}},\ }\href@noop {} {\bibfield
   {journal} {\bibinfo  {journal} {Nature Photonics}\ }\textbf {\bibinfo
  {volume} {12}},\ \bibinfo {pages} {659} (\bibinfo {year} {2018})}\BibitemShut
  {NoStop}%
\bibitem [{\citenamefont {Vercruysse}\ \emph {et~al.}(2019)\citenamefont
  {Vercruysse}, \citenamefont {Sapra}, \citenamefont {Su}, \citenamefont
  {Trivedi},\ and\ \citenamefont
  {Vu{\v{c}}kovi{\'c}}}]{vercruysse2019analytical}%
  \BibitemOpen
  \bibfield  {author} {\bibinfo {author} {\bibfnamefont {D.}~\bibnamefont
  {Vercruysse}}, \bibinfo {author} {\bibfnamefont {N.~V.}\ \bibnamefont
  {Sapra}}, \bibinfo {author} {\bibfnamefont {L.}~\bibnamefont {Su}}, \bibinfo
  {author} {\bibfnamefont {R.}~\bibnamefont {Trivedi}}, \ and\ \bibinfo
  {author} {\bibfnamefont {J.}~\bibnamefont {Vu{\v{c}}kovi{\'c}}},\ }\href@noop
  {} {\bibfield  {journal} {\bibinfo  {journal} {Scientific reports}\ }\textbf
  {\bibinfo {volume} {9}},\ \bibinfo {pages} {8999} (\bibinfo {year}
  {2019})}\BibitemShut {NoStop}%
\bibitem [{\citenamefont {Miller}\ and\ \citenamefont
  {Yablonovitch}(2015)}]{Miller2015}%
  \BibitemOpen
  \bibfield  {author} {\bibinfo {author} {\bibfnamefont {O.~D.}\ \bibnamefont
  {Miller}}\ and\ \bibinfo {author} {\bibfnamefont {E.}~\bibnamefont
  {Yablonovitch}},\ }\enquote {\bibinfo {title} {Inverse optical design},}\ in\
  \href {\doibase 10.1007/978-3-540-70529-1_45} {\emph {\bibinfo {booktitle}
  {Encyclopedia of Applied and Computational Mathematics}}},\ \bibinfo {editor}
  {edited by\ \bibinfo {editor} {\bibfnamefont {B.}~\bibnamefont {Engquist}}}\
  (\bibinfo  {publisher} {Springer Berlin Heidelberg},\ \bibinfo {address}
  {Berlin, Heidelberg},\ \bibinfo {year} {2015})\ pp.\ \bibinfo {pages}
  {729--732}\BibitemShut {NoStop}%
\bibitem [{\citenamefont {Offrein}\ \emph {et~al.}(1998)\citenamefont
  {Offrein}, \citenamefont {Bona}, \citenamefont {Germann}, \citenamefont
  {Massarek}, \citenamefont {Erni} \emph {et~al.}}]{offrein1998very}%
  \BibitemOpen
  \bibfield  {author} {\bibinfo {author} {\bibfnamefont {B.~J.}\ \bibnamefont
  {Offrein}}, \bibinfo {author} {\bibfnamefont {G.-L.}\ \bibnamefont {Bona}},
  \bibinfo {author} {\bibfnamefont {R.}~\bibnamefont {Germann}}, \bibinfo
  {author} {\bibfnamefont {I.}~\bibnamefont {Massarek}}, \bibinfo {author}
  {\bibfnamefont {D.}~\bibnamefont {Erni}},  \emph {et~al.},\ }\href@noop {}
  {\bibfield  {journal} {\bibinfo  {journal} {Journal of Lightwave Technology}\
  }\textbf {\bibinfo {volume} {16}},\ \bibinfo {pages} {1680} (\bibinfo {year}
  {1998})}\BibitemShut {NoStop}%
\bibitem [{\citenamefont {Mak}\ \emph {et~al.}(2016)\citenamefont {Mak},
  \citenamefont {Sideris}, \citenamefont {Jeong}, \citenamefont {Hajimiri},\
  and\ \citenamefont {Poon}}]{mak2016binary}%
  \BibitemOpen
  \bibfield  {author} {\bibinfo {author} {\bibfnamefont {J.~C.}\ \bibnamefont
  {Mak}}, \bibinfo {author} {\bibfnamefont {C.}~\bibnamefont {Sideris}},
  \bibinfo {author} {\bibfnamefont {J.}~\bibnamefont {Jeong}}, \bibinfo
  {author} {\bibfnamefont {A.}~\bibnamefont {Hajimiri}}, \ and\ \bibinfo
  {author} {\bibfnamefont {J.~K.}\ \bibnamefont {Poon}},\ }\href@noop {}
  {\bibfield  {journal} {\bibinfo  {journal} {Optics letters}\ }\textbf
  {\bibinfo {volume} {41}},\ \bibinfo {pages} {3868} (\bibinfo {year}
  {2016})}\BibitemShut {NoStop}%
\bibitem [{\citenamefont {Shen}\ \emph {et~al.}(2015)\citenamefont {Shen},
  \citenamefont {Wang}, \citenamefont {Polson},\ and\ \citenamefont
  {Menon}}]{shen2015integrated}%
  \BibitemOpen
  \bibfield  {author} {\bibinfo {author} {\bibfnamefont {B.}~\bibnamefont
  {Shen}}, \bibinfo {author} {\bibfnamefont {P.}~\bibnamefont {Wang}}, \bibinfo
  {author} {\bibfnamefont {R.}~\bibnamefont {Polson}}, \ and\ \bibinfo {author}
  {\bibfnamefont {R.}~\bibnamefont {Menon}},\ }\href@noop {} {\bibfield
  {journal} {\bibinfo  {journal} {Nature Photonics}\ }\textbf {\bibinfo
  {volume} {9}},\ \bibinfo {pages} {378} (\bibinfo {year} {2015})}\BibitemShut
  {NoStop}%
\bibitem [{\citenamefont {Zhou}\ \emph {et~al.}(2015)\citenamefont {Zhou},
  \citenamefont {Lazarov}, \citenamefont {Wang},\ and\ \citenamefont
  {Sigmund}}]{zhou2015minimum}%
  \BibitemOpen
  \bibfield  {author} {\bibinfo {author} {\bibfnamefont {M.}~\bibnamefont
  {Zhou}}, \bibinfo {author} {\bibfnamefont {B.~S.}\ \bibnamefont {Lazarov}},
  \bibinfo {author} {\bibfnamefont {F.}~\bibnamefont {Wang}}, \ and\ \bibinfo
  {author} {\bibfnamefont {O.}~\bibnamefont {Sigmund}},\ }\href@noop {}
  {\bibfield  {journal} {\bibinfo  {journal} {Computer Methods in Applied
  Mechanics and Engineering}\ }\textbf {\bibinfo {volume} {293}},\ \bibinfo
  {pages} {266} (\bibinfo {year} {2015})}\BibitemShut {NoStop}%
\bibitem [{\citenamefont {Su}\ \emph {et~al.}(2017)\citenamefont {Su},
  \citenamefont {Piggott}, \citenamefont {Sapra}, \citenamefont {Petykiewicz},\
  and\ \citenamefont {Vuckovic}}]{su2017inverse}%
  \BibitemOpen
  \bibfield  {author} {\bibinfo {author} {\bibfnamefont {L.}~\bibnamefont
  {Su}}, \bibinfo {author} {\bibfnamefont {A.~Y.}\ \bibnamefont {Piggott}},
  \bibinfo {author} {\bibfnamefont {N.~V.}\ \bibnamefont {Sapra}}, \bibinfo
  {author} {\bibfnamefont {J.}~\bibnamefont {Petykiewicz}}, \ and\ \bibinfo
  {author} {\bibfnamefont {J.}~\bibnamefont {Vuckovic}},\ }\href@noop {}
  {\bibfield  {journal} {\bibinfo  {journal} {ACS Photonics}\ }\textbf
  {\bibinfo {volume} {5}},\ \bibinfo {pages} {301} (\bibinfo {year}
  {2017})}\BibitemShut {NoStop}%
\bibitem [{\citenamefont {Qian}(2013)}]{qian2013topology}%
  \BibitemOpen
  \bibfield  {author} {\bibinfo {author} {\bibfnamefont {X.}~\bibnamefont
  {Qian}},\ }\href@noop {} {\bibfield  {journal} {\bibinfo  {journal} {Computer
  Methods in Applied Mechanics and Engineering}\ }\textbf {\bibinfo {volume}
  {265}},\ \bibinfo {pages} {15} (\bibinfo {year} {2013})}\BibitemShut
  {NoStop}%
\bibitem [{\citenamefont {De~Boor}(1972)}]{de1972calculating}%
  \BibitemOpen
  \bibfield  {author} {\bibinfo {author} {\bibfnamefont {C.}~\bibnamefont
  {De~Boor}},\ }\href@noop {} {\bibfield  {journal} {\bibinfo  {journal}
  {Journal of Approximation theory}\ }\textbf {\bibinfo {volume} {6}},\
  \bibinfo {pages} {50} (\bibinfo {year} {1972})}\BibitemShut {NoStop}%
\bibitem [{\citenamefont {Cox}(1972)}]{cox1972numerical}%
  \BibitemOpen
  \bibfield  {author} {\bibinfo {author} {\bibfnamefont {M.~G.}\ \bibnamefont
  {Cox}},\ }\href@noop {} {\bibfield  {journal} {\bibinfo  {journal} {IMA
  Journal of Applied Mathematics}\ }\textbf {\bibinfo {volume} {10}},\ \bibinfo
  {pages} {134} (\bibinfo {year} {1972})}\BibitemShut {NoStop}%
\bibitem [{\citenamefont {Kroon}(2011)}]{ISOCONTOUR}%
  \BibitemOpen
  \bibfield  {author} {\bibinfo {author} {\bibfnamefont {D.~J.}\ \bibnamefont
  {Kroon}},\ }\href@noop {} {\enquote {\bibinfo {title} {{Isocontour}
  {W}ebpage},}\ } (\bibinfo {year} {2011}),\ \bibinfo {note}
  {{https://www.mathworks.com/matlabcentral/fileexchange
  /30525-isocontour}}\BibitemShut {NoStop}%
\bibitem [{\citenamefont {Bernard}\ \emph {et~al.}(2009)\citenamefont
  {Bernard}, \citenamefont {Friboulet}, \citenamefont {Th{\'e}venaz},\ and\
  \citenamefont {Unser}}]{bernard2009variational}%
  \BibitemOpen
  \bibfield  {author} {\bibinfo {author} {\bibfnamefont {O.}~\bibnamefont
  {Bernard}}, \bibinfo {author} {\bibfnamefont {D.}~\bibnamefont {Friboulet}},
  \bibinfo {author} {\bibfnamefont {P.}~\bibnamefont {Th{\'e}venaz}}, \ and\
  \bibinfo {author} {\bibfnamefont {M.}~\bibnamefont {Unser}},\ }\href@noop {}
  {\bibfield  {journal} {\bibinfo  {journal} {IEEE Transactions on Image
  Processing}\ }\textbf {\bibinfo {volume} {18}},\ \bibinfo {pages} {1179}
  (\bibinfo {year} {2009})}\BibitemShut {NoStop}%
\bibitem [{\citenamefont {Khoram}\ \emph {et~al.}(2019)\citenamefont {Khoram},
  \citenamefont {Chen}, \citenamefont {Liu}, \citenamefont {Ying},
  \citenamefont {Wang}, \citenamefont {Yuan},\ and\ \citenamefont
  {Yu}}]{khoram2019nanophotonic}%
  \BibitemOpen
  \bibfield  {author} {\bibinfo {author} {\bibfnamefont {E.}~\bibnamefont
  {Khoram}}, \bibinfo {author} {\bibfnamefont {A.}~\bibnamefont {Chen}},
  \bibinfo {author} {\bibfnamefont {D.}~\bibnamefont {Liu}}, \bibinfo {author}
  {\bibfnamefont {L.}~\bibnamefont {Ying}}, \bibinfo {author} {\bibfnamefont
  {Q.}~\bibnamefont {Wang}}, \bibinfo {author} {\bibfnamefont {M.}~\bibnamefont
  {Yuan}}, \ and\ \bibinfo {author} {\bibfnamefont {Z.}~\bibnamefont {Yu}},\
  }\href@noop {} {\bibfield  {journal} {\bibinfo  {journal} {Photonics
  Research}\ }\textbf {\bibinfo {volume} {7}},\ \bibinfo {pages} {823}
  (\bibinfo {year} {2019})}\BibitemShut {NoStop}%
\end{thebibliography}%

\end{document}